\documentclass[twocolumn,prb,showpacs,preprintnumbers,amsmath,amssymb]{revtex4}

\usepackage{graphicx}
\usepackage{dcolumn}
\usepackage{bm}
\usepackage{epsfig}

\begin{document}


\title{A new interpretation of dielectric data in molecular glass formers}

\author{U. Buchenau}
\email{buchenau-juelich@t-online.de}
\author{M. Ohl}
\author{A. Wischnewski}
\affiliation{%
Institut f\"ur Festk\"orperforschung, Forschungszentrum J\"ulich\\
Postfach 1913, D--52425 J\"ulich, Federal Republic of Germany
}%
\date{Januar 13, 2006}

\begin{abstract}
Literature dielectric data of glycerol, propylene carbonate and 
ortho-terphenyl (OTP)
show that the measured dielectric relaxation is a decade faster
than the Debye expectation, but still a decade slower than the
breakdown of the shear modulus. From a comparison of time scales,
the dielectric relaxation seems to be due to a process which
relaxes not only the molecular orientation, but the entropy, the
short-range order and the density as well. On the basis of this
finding, we propose an alternative to the Gemant-DiMarzio-Bishop
extension of the Debye picture.
\end{abstract}


\maketitle

Broadband dielectric spectroscopy has developed into the most
important tool for the study of glass formers. It is able to cover
the whole relevant frequency range, from $\mu$Hz to THz
\cite{loidl}. Therefore it would be very desirable to understand
the dielectric susceptibility in terms of physical processes. In
particular, one would like to link the $\alpha$-peak of the
dielectric data to the disappearance of the shear modulus at long
times, the essence of the flow process.

Such a link is in principle provided by the Gemant-DiMarzio-Bishop
\cite{gemant,dimarzio} extension (GDB extension) of Debye's
treatment. The extension considers the molecule as a small sphere
with a hydrodynamic radius $r_H$ immersed in the viscoelastic liquid.
The medium is characterized by a frequency-dependent complex shear
modulus
\begin{equation}\label{gom}
	G(\omega)=G_\infty g(\omega),
\end{equation}
where $G_\infty$ is the infinite-frequency shear modulus and $g(\omega)$
is a normalized complex function, increasing from zero to one as the frequency
goes from zero to infinity.

For molecules with a weak dipole moment like OTP, the GDB extension is
\begin{equation}\label{gdb}
\frac{\epsilon(\omega)-n^2}{\epsilon_{low}-n^2}=\frac{1}{1+c_rg(\omega)}
\end{equation}
with
\begin{equation}\label{cr}
c_r=\frac{4\pi G_\infty r_H^3}{k_BT}.
\end{equation}
Here $\epsilon(\omega)$ is the complex frequency-dependent
dielectric constant (with the conductivity contribution already
subtracted), $n$ is the refractive index, $\epsilon_{low}$ is the
low-frequency limit of $\epsilon$ and $T$ is the
temperature. One needs only the knowledge of the molecular radius
$r_H$. Then one can calculate $\epsilon$ from measurable quantities.

For strongly polar molecules like glycerol and propylene carbonate,
one should take the difference between the external applied electric
field and the internal field seen by the molecule into account 
\cite{boettcher}. In Onsager's scheme \cite{onsager}, extended to
dynamics by Fatuzzo and Mason \cite{fatuzzo} and reformulated by Niss
and Jakobsen \cite{niss1} 
\begin{equation}\label{fat}
\frac{\epsilon(\omega)-n^2}{\epsilon_{low}-n^2}
\frac{2\epsilon_l\epsilon(\omega)+\epsilon_ln^2}
{\epsilon^2(\omega)+\epsilon(\omega)\epsilon_l+\epsilon_ln^2}.
=\frac{1}{1+c_rg(\omega)}
\end{equation}
This complex quadratic equation still allows to calculate $\epsilon(\omega)$
from $G(\omega)$, provided $r_H$ is known.

\begin{figure}[h]
	\centering
		\includegraphics{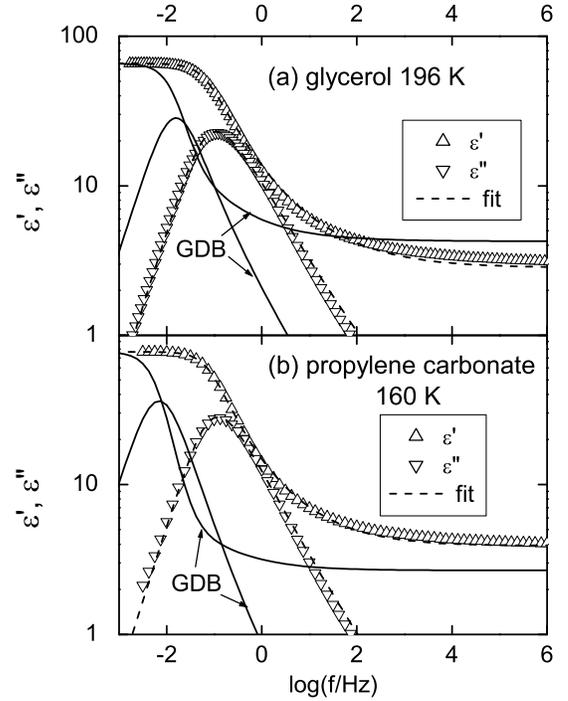}
	\label{fig:fig1}
 \caption{Gemant-DiMarzio-Bishop (GDB) expectation (continuous
lines) compared to (a) dielectric data \cite{rossler} of glycerol
at 196 K (b) dielectric data \cite{loidl2} of propylene carbonate
at 160 K . The better fits (dashed lines) are explained in the
text.}
\end{figure}

The molecular radius can be determined from NMR field gradient
diffusion data for glycerol  \cite{chang} ($r_H=0.16 nm$) and for
propylene carbonate \cite{qi} ($r_H=0.26 nm$) via the Stokes-Einstein
equation. For glycerol, there is a dynamic shear modulus
measurement \cite{donth1} in the temperature range close to the
glass transition. Using the shift factors of this measurement, one
can calculate $G(\omega)$ at the temperature 196 K of a dielectric
measurement \cite{rossler}. At this temperature \cite{comez1}, $n^2=2.26$.
Fig. 1 (a) compares the GDB expectation of eq. (\ref{fat}) with these
values to measured data. The calculation underestimates the peak
frequency in $\epsilon''$ by an order of magnitude, a
discrepancy which has been noted earlier \cite{chang}.

The same holds for propylene carbonate. There is a shear modulus
measurement \cite{donth2} at 159 K, close to the glass transition.
Fig. 1 (b) compares the GDB prediction (with $n^2=2.19$
\cite{brodin}) to dielectric data \cite{loidl2} at 160 K. Again,
we find the peak in $\epsilon''$ shifted by a decade.
Obviously, the undercooled liquid finds a much faster way to relax
the molecular orientation than the Debye mechanism of a molecular
sphere rotating in a viscous liquid.

As pointed out by Niss, Jakobsen and Olsen \cite{niss}, one does not even
get a good fit if one adapts the molecular radius, because if one
adapts the peak in $\epsilon''$, the high-frequency $\epsilon'$
gets much too high.

In order to find out which mechanism might be responsible for the
decay of the molecular orientation, we compare the dielectric
relaxation times to the ones determined by other
techniques, in the spirit of earlier comparisons by Ngai and
Rendell \cite{ngai}, Blochowicz et al \cite{bloch} and by
Schr\"oter and Donth \cite{donth2}. We recalculate all data in
terms of a Kohlrausch-Williams-Watts (KWW) decay in time
exp$(-(t/\tau_{KWW})^\beta)$, either by using pragmatical recipes from
the literature \cite{patt,alvarez} or by refitting the data. To get rid
of the strong temperature dependence of the
viscosity, the resulting KWW relaxation time is divided by the Maxwell time
$\tau_{Maxwell}=\eta/G_\infty$.

The choice of a Kohlrausch or KWW-function is practical for the following reasons:

(i) it very often gives a good fit

(ii) the inverse of the absorption peak frequency in $\omega$ is close to $\tau_{KWW}$, so one
compares peak frequencies, independent of the stretching parameter $\beta$

(iii) for a shear modulus following the Kohlrausch function
\begin{equation}\label{taukww}
\frac{\tau_{KWW}}{\tau_{Maxwell}}=\frac{\beta}{\Gamma(1/\beta)}.	
\end{equation}
Usually, $\beta$ lies between 0.4 and 0.6, so the ratio should
be between one third and two thirds; the Kohlrausch relaxation time should be a factor 1.5 to 3 shorter
than the Maxwell time

(iv) for a shear compliance measurement like the one \cite{plazek} in OTP, the steady-state
compliance $J^0_e$ obeys the relation \cite{ferry}
\begin{equation}\label{j0ef}
J^0_e\eta^2=\int_0^\infty tG(t)dt,	
\end{equation}
so for a Kohlrausch function \cite{alvarez}
\begin{equation}\label{j0e}
J^0_eG_\infty=\frac{\Gamma(2/\beta)\beta}{\Gamma((1/\beta)}.
\end{equation}
Since the compliance measurement supplies all three values $J^0_e$, $\eta$ and $G_\infty$,
one can determine $\beta$ and $\tau_{KWW}/\tau_{Maxwell}$ without calculating $\tau_{Maxwell}$.

As usual, the measured viscosity $\eta$ of our
three substances is fitted in terms of a combination of two
Vogel-Fulcher-Tammann-Hesse laws
\begin{equation}\label{vfth}
\log\eta=\log\eta_{0i}+\frac{B_i}{T-T_{0i}}	
\end{equation}
with $i=1$ and $i=2$, respectively.
The first of these two is valid below a temperature $T_1$, the second
above a temperature $T_2\leq T_1$. Between $T_2$ and $T_1$, one takes
a linear interpolation between the two to ensure continuity.

\begin{table}[h]
	\centering
		\begin{tabular}{|c|c|c|c|}
\hline
  substance &   glycerol   & propylene carbonate &      OTP     \\
\hline
log($\eta_{01}$/Pa s) & -7.1  & -8.92    & -11.89    \\			
$B_1$ (K) & 1260   &  667   & 1461.2    \\			
$T_{01}$ (K) & 118  & 122    & 178.4    \\			
$T_1$ (K) & 283  &  193   & 310    \\			
log($\eta_{02}$/Pa s) & -5.45  & -3.91    & -4.24    \\			
$B_2$ (K) & 780  & 191    &  245.9   \\			
$T_{02}$ (K) & 153  & 150    & 241.72    \\			
$T_2$ (K) & 283  &  175   & 275    \\			
$T_g$ (K) & 187  &  157   & 243   \\			
$G_\infty(T_g)$ (GPa) & 4.58  & 2.5    &  1.6   \\			
$a$ (K$^{-1}$) & 0.023  &  0.007   &  0.0057   \\			
$b$ (K$^{-2}$) & 2.1$\cdot 10^{-5}$  & 2$\cdot 10^{-5}$    &  1.4$\cdot 10^{-5}$   \\	
\hline		
		\end{tabular}
	\caption{Viscosity and shear modulus parameters. References see text.}
	\label{tab1}
\end{table}

The infinity frequency shear modulus $G_\infty$ is parametrized
in terms of a Taylor expansion around the glass temperature $T_g$
\begin{equation}\label{ginf}
G_\infty=G_\infty(T_g)(1-a(T-T_g)+b(T-T_g)^2).
\end{equation}

The parameters of these two equations for the viscosity and the
infinite frequency shear modulus are listed in Table I. For
glycerol \cite{donth1} and propylene carbonate \cite{bondeau},
the Vogel-Fulcher parameters were taken from viscosity data fits
in the literature. For OTP, we fitted our own parameters to the many
viscosity measurements in the literature 
\cite{greet1,greet2,uhl1,uhl2,plazek}. In glycerol, $G_\infty$ was fitted to the
Brillouin shear wave measurement of Scarponi {\it et al} \cite{comez2}. For
propylene carbonate, there is no Brillouin shear wave measurement. 
Therefore, the infinite frequency shear modulus had to be taken
from a longitudinal Brillouin
measurement \cite{lars}, assuming $G_\infty=c_{11}/4$. In OTP, there is
a shear wave Brillouin scattering measurement \cite{dreyf}, but the
$G_\infty$-values from this measurement extrapolate to zero already at
308 K. Therefore we took $G_\infty(T_g)$ from this measurement, but
determined the parameters $a$ and $b$ of eq. (\ref{ginf}) from the
combined evaluation of light and x-ray Brillouin scattering of Monaco
{\it et al} \cite{monaco}, assuming the same temperature dependence
for the infinite frequency longitudinal and shear moduli.

We start the comparison for glycerol in Fig. 2. The normalized dielectric
\cite{loidl,stickel,rossler} KWW relaxation times are compared to those from
mechanical data
\cite{donth1,donth2,picci,jeong1,olsen1,jeong2,nelson,comez1},
from dynamic heat capacity measurements \cite{birge,korus}, from
NMR \cite{reinsberg}, from PCS (photon correlation spectroscopy)
\cite{dux}, from TG (transient grating) \cite{nelson} and from
neutron spin-echo measurements at the first sharp diffraction peak
\cite{wuttke}. Dynamic light scattering (DLS) data \cite{brodin2}
(not shown in Fig. 2) tend to lie between dielectrics and
mechanics, but otherwise the figure corroborates the earlier
conclusion of Schr\"oter and Donth \cite{donth2}, namely that
there seems to be a grouping into the faster mechanical relaxation
and a slower heat capacity, dielectric, NMR, PCS, TG and neutron
spin-echo relaxation. The mechanical relaxation times follow the
expectation of eq. (\ref{taukww}) within reasonable error limits. The others
tend to lie a factor of about ten higher.
With changing temperature, both time scales
move together with a roughly constant separation. This shows that
the misfit of the Debye result is temperature-independent, unlike
the deviations from the Stokes-Einstein relation at lower
temperatures \cite{chang}. We will come back to this point
in the discussion of OTP. Here, let us first discuss what one sees
in each technique.

\begin{figure}[h]
	\centering
		\includegraphics[width=0.50\textwidth]{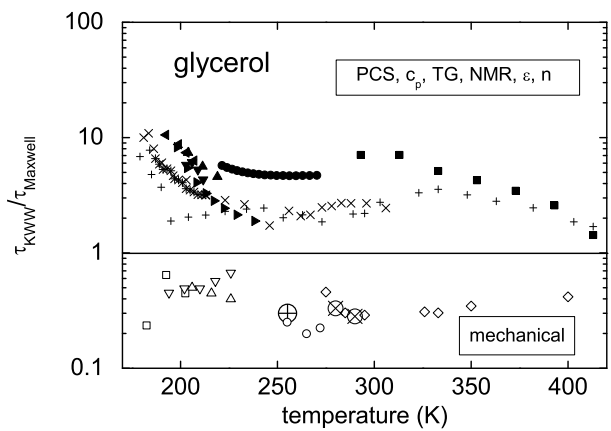}
	\label{fig:fig2}
	\caption{Kohlrausch-Williams-Watts relaxation times in glycerol,
normalized to the Maxwell time as described in the text. Symbols:
open squares shear \cite{donth2}; open up triangles shear
\cite{jeong1}; open down triangles compression \cite{olsen1}; open
circles longitudinal acoustic \cite{jeong2}; open circle with plus
shear and compression \cite{picci}; open circles with cross TG
\cite{nelson}; open diamonds longitudinal Brillouin \cite{comez1};
plusses dielectric \cite{loidl}; crosses dielectric
\cite{rossler}; asterisks dielectric \cite{stickel}; full up
triangles heat capacity \cite{birge}; full down triangles heat
capacity \cite{korus}; full left triangles NMR \cite{reinsberg};
full right triangles PCS \cite{dux}; full circles TG
\cite{nelson}; full squares neutron spin-echo \cite{wuttke}.}
\end{figure}

In the case of mechanical and dielectric data, there can be large
differences in relaxation time between a modulus and the
corresponding compliance. This difference is negligible if the
relative change of the quantity in question is small, but here we
deal with large relative changes. Therefore one has to check
whether mechanical moduli and dielectric susceptibility are the
correct choice.

For the mechanical data, there is a good physical reason to choose
the moduli rather than the compliances, because this choice leads
to practically the same relaxation time for the bulk and the shear
modulus \cite{picci,meister,olsen2}. In fact, glycerol was one of
the first cases in which this equality was demonstrated by the
longitudinal and shear ultrasonic data of Piccirelli and Litovitz
\cite{picci} (the circle with a plus in Fig. 2 at 255 K). In Fig.
2, it is again demonstrated at lower frequencies by the good
agreement between the shear measurements
\cite{donth1,donth2,jeong1} and the compression measurement of
Christensen and Olsen \cite{olsen1}. If one goes over to
compliances, this good agreement gets lost \cite{foot}.

Similarly, in the dielectric case one should take the dielectric
constant rather than its inverse. Otherwise, the good agreement
between NMR and dielectric constant \cite{qi,bloch} (which is
natural because both techniques sample the molecular orientation)
would get completely lost.

The transient grating (TG) experiment \cite{nelson} measures both
the damping of longitudinal sound waves and a longer structural
relaxation time \cite{yn,dreyfus}. One does not get the
longitudinal sound wave relaxation time directly, but one can
extract it from the temperature dependence of the damping. With a
fitted $\beta=0.5$, the damping of the sound waves translates into
the two circles with crosses at 280 and 290 K in Fig. 2. In this
case, one sees the splitting of time scales within experiments
with a single sample and with the same temperature sensor.

In the structural relaxation of the transient grating experiment,
the heat of the phonon bath transforms into structural potential energy,
thereby expanding the sample. The relaxation time of this process
is intimately related to the relaxation times of the heat
capacity and of the density, which in turn are related to
each other. The latter relation has been discussed earlier in several papers
\cite{latz,jaeckle,dyre,christensen,simon}. Photon correlation spectroscopy (PCS)
measures the refraction index fluctuations on the scale of the wavelength of
the light, essentially density fluctuations. So it is not surprising
to find the structural TG relaxation times close to those of PCS and
heat capacity. The neutron spin-echo measurements at the first
sharp diffraction peak sample the decay of the short-range order
of the molecular array. Again, it is not unexpected to find them close to those
of the density and the entropy. What is surprising is to find the dielectric
and the NMR times in the same group, because we are used to think of them
as a single-molecule property \cite{chang} and not as a collective process.

The idea of two different time scales (or an initial and final
stage of the same process) is further supported by the different
shape of the relaxation functions for the two groups of Fig. 2.
The mechanical data have a decidedly larger stretching
($\beta_{KWW}\approx0.4..0.5$) than the heat capacity, the
dielectric constant and the neutron spin-echo signal
($\beta_{KWW}\approx0.55..0.7$).

The same splitting of time scales, though not for so many
different techniques as in the heavily studied case of glycerol,
has been found for propylene carbonate \cite{brodin} and has been
discussed in the framework of the mode coupling theory \cite{mct}.
Note that this time scale splitting is not the two-stage scenario
of the mode-coupling theory, because both time scales move
together with the Maxwell time. In fact, in ref. \cite{brodin} the
$\alpha$-process of the theory was not attributed to the slower,
but to the faster process.

\begin{figure}[h]
	\centering
		\includegraphics[width=0.50\textwidth]{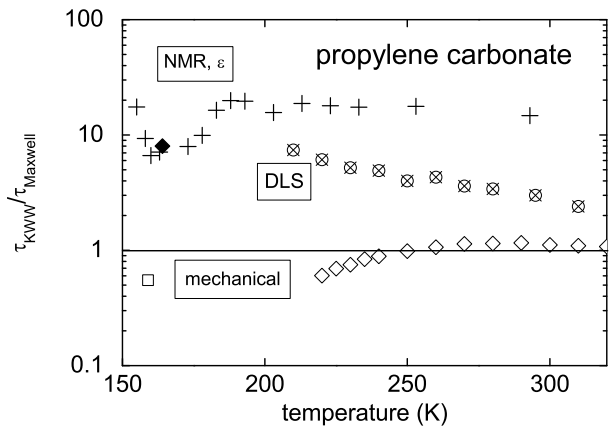}
	\label{fig:fig3}
 \caption{Kohlrausch-Williams-Watts relaxation times in propylene
carbonate, normalized to the Maxwell time as described in the
text. Symbols: open square shear \cite{donth2}; open diamonds
longitudinal Brillouin \cite{brodin}; plusses dielectric
\cite{loidl2}; full diamond NMR \cite{qi}; open circles with cross
DLS \cite{du}.}	
\end{figure}

With the parameters in Table I, one can again relate
the measured Kohlrausch relaxation times to the Maxwell time.
As in glycerol, mechanical shear
\cite{donth2} and Brillouin \cite{brodin} data in Fig. 3 show a
decade faster decay than NMR \cite{qi} and dielectric
\cite{loidl2} measurements, while DLS data \cite{du} lie in
between.

Fig. 4 shows again a heavily studied case, OTP. Mechanical measurements
include a shear compliance study \cite{plazek}, longitudinal ultrasonic
data \cite{monaco}, a transient grating experiment \cite{torre},
a transverse Brillouin measurement \cite{dreyf} and a thorough analysis of
longitudinal Brillouin light and x-ray scattering \cite{monaco}. Of these,
the shear compliance, the ultrasonic and the longitudinal Brillouin
data follow the Kohlrausch expectation of eq. (\ref{taukww}), but the
transverse Brillouin data and the longitudinal sound wave part of the 
transient grating results do not; they
show a sudden rise at about 290 K. The reason for this deviation is clearly revealed
in the analysis of Monaco {\it et al} \cite{monaco}. At 290 K, the Johari-
Goldstein peak merges with the main $\alpha$-relaxation. At such a point,
our analysis in terms of a single Kohlrausch function is bound to fail.

Otherwise, Fig. 4 corroborates the results in Fig. 2. Again, the structural
relaxation time from the TG experiment \cite{torre}, heat capacity \cite{leyser},
PCS \cite{fytas,fischer}
and NMR \cite{chang2} lie close to the dielectric data \cite{hansen}. The neutron
data at the first sharp diffraction peak \cite{toelle} lie a bit lower, but
do still clearly belong to the upper group. The dynamic light
scattering points \cite{fischer,cummins} do not lie between the two groups
as in glycerol and propylene carbonate,
but have higher relaxation times than all the other experiments.

In OTP, there is a rather convincing explanation of the NMR data
in terms of a single-molecule picture \cite{chang2}, describing them in terms of
rotational diffusion which follows the Debye-Stokes-Einstein equation
\begin{equation}
D_{trans}=\frac{k_BT}{6\pi\eta r_H}=\frac{4}{3}r_H^2D_{rot},
\end{equation}
where $D_{trans}$ is the translational diffusion constant of the molecule and
$D_{rot}$ is its rotational diffusion constant. For continuous rotational diffusion, the
relaxation time for the Legendre polynomials is
\begin{equation}
	\tau_{L,rot}=\frac{1}{L(L+1)D_{rot}},
\end{equation}
where $L$ is the order of the Legendre polynomial. For the dielectric signal,
$L=1$, but the NMR measurements referenced
so far are all two-pulse sequences for deuterated molecules, where $L=2$. The
OTP data \cite{chang2} are well described
with a hydrodynamic radius $r_H$ of 22 nm, close to the value $r_H$ = 23 nm found
in NMR field gradient measurements \cite{chang2} at temperatures above 1.2 $T_g$.
The values are smaller than the expected van-der-Waals radius of 37 nm, but still
not too far away from it.

\begin{figure}[h]
	\centering
		\includegraphics[width=0.50\textwidth]{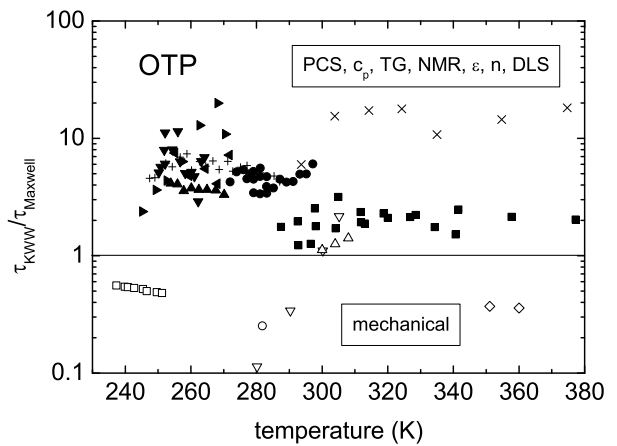}
	\label{fig:fig4}
 \caption{Kohlrausch-Williams-Watts relaxation times in OTP,
normalized to the Maxwell time as described in the text. Symbols:
open squares shear \cite{plazek}; open
circles longitudinal acoustic \cite{monaco}; open up triangles TG
\cite{torre}; open down triangles transverse Brillouin \cite{dreyf};
open diamonds longitudinal Brillouin \cite{monaco};
plusses dielectric \cite{hansen}; full down triangles NMR
\cite{chang2}; full up
triangles heat capacity \cite{leyser}; full right triangles PCS
\cite{fischer}; full left triangles PCS \cite{fytas}; full circles TG
\cite{torre}; full squares neutron spin-echo \cite{toelle};
crosses DLS \cite{fischer,cummins}.}
\end{figure}

If one
lowers the temperature, the rotational relaxation time follows the temperature
dependence of the viscosity, while the translational diffusion begins to deviate
towards higher values. The same decoupling between translational and rotational
motion has been found in photobleaching experiments \cite{ediger} with guest
molecules in OTP and has been taken as evidence for dynamical heterogeneity.
In these experiments, one observes an increase of the relaxation times with
increasing molecular diameter as expected, giving additional support to the
single-molecule concept.

But the GDB extension of this single-molecule picture
to describe the relation between $G(\omega)$ and $\epsilon(\omega)$,
eq. (\ref{gdb}), does not work. At the glass temperature of OTP with $G_\infty=1.6$ GPa, 
one calculates a $c_r$ of 80 from eq. (\ref{cr}). This implies that the peak in
$\epsilon''(\omega)$ should be at a factor of 80 lower frequency than the one in
$G''(\omega)$, while the experiment shows only a factor of ten. Again, this discrepancy
has been noted before \cite{chang}. In this case, one cannot
blame the difference between external and internal electric field, because the dipole
moment of OTP is very small.

In some cases, one even finds the peak in $\epsilon''(\omega)$ rather close to the one
in $G''(\omega)$. In decahydroisoquinoline (DHIQ) at $T_g$, they lie only a factor of 1.6
apart \cite{niss}, instead of the factor of about 100 that one expects.

Also, the single-molecule picture fails to explain the striking coincidence
between dielectric and NMR relaxation times on the one hand and heat
capacity, density and short range order relaxation times on the other.

We will pursue an alternative explanation for the data in Figs. 2 to 4,
namely that the flow or $\alpha$-process begins at short times with a
breakdown of the mechanical rigidity (the lower half of points in
Figs. 2, 3 and 4). A decade in time later, there seems to be a final
process which equilibrates everything, the density, the entropy
and the short range order (the upper half of points in the three
figures). This final process equilibrates also the molecular
orientation, an order of magnitude earlier in time than expected
on the basis of the Debye concept. In fact, a recent aging
experiment \cite{lunkenheimer} shows that the dielectric
relaxation time is indeed the final aging relaxation time also in
a number of other molecular glass formers.

A possible way to understand such a process is to postulate a
configurational potential energy which has a small fraction of shear energy,
able to decay within the mechanical relaxation time, while the
large rest is merely feeding the shear energy. In a physical
picture, one divides the potential energy of a given configuration
into a long range shear part and everything else. This "everything
else" is supposed to be harmonic, decaying only via the shear
energy channel.

This is similar to the physical mechanism of the Debye process,
where the feeding energy is the energy of the electric dipole in
the electric field. For the configurational energy, one replaces
the electric dipole energy by the mechanical energy of an harmonic
oscillator. As we will see, this change leads to a different
equation; the two cases are similar, but not identical.

To formulate the concept quantitatively, let us consider a
mechanical model, a small spring $r$ in series with a
frequency-dependent spring $g(\omega)=G(\omega)/G_\infty$. The
compliance of the combination is the sum of the two compliances.
Normalizing this sum, one gets
\begin{equation}\label{inv}
\Phi(\omega)=\frac{1+r}{1+r/g(\omega)}.
\end{equation}
This is the Fourier transform of the decay function of the
configurational energy according to the postulate above.
$\Phi(\omega)$ is 1 in the high-frequency limit and zero in the
low-frequency limit; it is a modulus function.

We further postulate that the decay of the configuration involves
a complete reorientation of the molecules, so it is mirrored in
the dielectric signal. In the dielectric susceptibility, one
expects to see $1-\Phi(\omega)$. Then
\begin{equation}\label{epsfit}
\frac{\epsilon(\omega)-n^2}{\epsilon_{low}-n^2}=\frac{1-g(\omega)}{1+g(\omega)/r}.
\end{equation}
This relation differs from the Gemant-DiMarzio-Bishop relation,
eq. (\ref{gdb}), by the $1-g(\omega)$ in the numerator.

We used eq. (\ref{epsfit}) to fit the dielectric data of glycerol
\cite{loidl,rossler} and propylene carbonate \cite{loidl2}.
$g(\omega)$ was obtained by first fitting the dynamical shear data
\cite{donth1,donth2} at the glass transition in terms of a KWW
function and then shifting this function to the required
temperature using the shift factors of the Maxwell time. In
glycerol, we also took the slight change of the Kohlrausch-$\beta$
of the shear with temperature into account.

To get a good fit, it turned out to be necessary to leave $n$ as a
free parameter and to allow for a slight difference of the shift
factor (remember that the shear and dielectric data stem from
different laboratories). These deviations, however, remained small
and of the order of the differences between the fit values for the
two different dielectric glycerol experiments
\cite{loidl,rossler}. For the glycerol fit shown in Fig. 1 (a),
the shift factor difference corresponded to a factor 0.7 in
relaxation time. The $n$-value was 1.67 instead of 1.50. For the
propylene carbonate fit in Fig. 1 (b), the shift factor difference
corresponded to a factor of 0.6 in relaxation time, and $n$ was
1.99 instead of 1.48. The shear energy fraction $r$ was 0.19 in
both fits. At higher temperatures, $r$ tended to diminish, more
strongly in propylene carbonate than in glycerol.

In glycerol, the so-called "excess wing" at higher frequencies has
been extensively discussed \cite{lunk,ngai2,ngai3}. From aging
\cite{lunk,ngai2} and pressure experiments \cite{ngai3}, one forms
the impression that the excess wing is an unresolved secondary
relaxation or Johari-Goldstein peak. Eq. (\ref{epsfit}) does not
add to this evidence, but demonstrates that one deals with both a
dielectric and a mechanical excess wing. In fact, once the
parameters are known, one can use the equation to calculate the
expected shear response from dielectric data. It remains to be
seen, however, how far one can trust an implicit assumption of eq.
(\ref{epsfit}), namely that the elementary relaxation processes
behind the breakdown of the shear modulus have the same ratio of
the mechanical to the dielectric dipole moment over the whole
frequency range.

To conclude, we propose an alternative to the Debye concept,
because the Debye scheme is unable to account for
dielectric data in glycerol, propylene carbonate and ortho-terphenyl.
The alternative is based on the finding that the
dielectric relaxation time in glycerol and OTP is close to the ones for
the density, the entropy and the short range order. The scheme
provides a good fit for broadband dielectric spectra of glycerol and propylene carbonate.

We thank Peter Lunkenheimer and Ernst R\"ossler for supplying
their dielectric data and for helpful discussions.


\begin{thebibliography}{99}
\bibitem{foot} One can in fact get a reasonable agreement between
longitudinal ultrasonic relaxation times and dielectric ones by
evaluating the ultrasonic data in terms of a compliance
\cite{jeong1}. But then one does not know what to do with the
shear data.
\end{thebibliography}
\end{document}